\documentclass[aps,prd,preprint,superscriptaddress,showpacs]{revtex4}

\usepackage{amsfonts}
\usepackage{mathrsfs}
\usepackage{amssymb}
\usepackage{amsmath}
\usepackage{graphicx,epsfig}

\def\bwt{\begin{widetext}}

\def\ewt{\end{widetext}}

\def\be{\begin{equation}}

\def\ee{\end{equation}}

\def\bea{\begin{eqnarray}}

\def\eea{\end{eqnarray}}

\def\bean{\begin{eqnarray*}}

\def\eean{\end{eqnarray*}}

\def\bary{\begin{array}}

\def\eary{\end{array}}

\def\bit{\begin{itemize}}

\def\eit{\end{itemize}}

\def\su5u1{SU(5) \times U(1)}

\def\fsu5u1{SU(5) \times U(1)'}

\def\so10{SO(10)}

\def\sq20{SO(10) \times SO(10)}

\usepackage{amssymb}

\begin{document}

\setlength{\parskip}{0cm}

\title{Dark Matter, Proton Decay and \\
Other Phenomenological Constraints in ${\cal F}$-$SU(5)$}

\author{Tianjun Li}

\affiliation{George P. and Cynthia W. Mitchell Institute for
Fundamental Physics, Texas A$\&$M University, College Station, TX
77843, USA }

\affiliation{Key Laboratory of Frontiers in Theoretical Physics,
      Institute of Theoretical Physics, Chinese Academy of Sciences,
Beijing 100190, P. R. China }

\author{James A. Maxin}

\affiliation{George P. and Cynthia W. Mitchell Institute for
Fundamental Physics, Texas A$\&$M University, College Station, TX
77843, USA }

\author{Dimitri V. Nanopoulos}

\affiliation{George P. and Cynthia W. Mitchell Institute for
Fundamental Physics,
 Texas A$\&$M University, College Station, TX 77843, USA }

\affiliation{Astroparticle Physics Group,
Houston Advanced Research Center (HARC),
Mitchell Campus, Woodlands, TX 77381, USA}

\affiliation{Academy of Athens, Division of Natural Sciences,
 28 Panepistimiou Avenue, Athens 10679, Greece }

\author{Joel W. Walker}

\affiliation{Department of Physics, Sam Houston State University,
Huntsville, TX 77341, USA }



\begin{abstract}

We study gravity mediated supersymmetry breaking in ${\cal F}$-$SU(5)$ and its low-energy supersymmetric phenomenology.
The gaugino masses are not unified at the traditional grand unification scale, but we
nonetheless have the same one-loop gaugino mass relation at the electroweak scale as minimal supergravity (mSUGRA).
We introduce parameters testable at the colliders to measure the small second loop deviation from the
mSUGRA gaugino mass relation at the electroweak scale. In the minimal $SU(5)$ model with gravity mediated
supersymmetry breaking, we show that the deviations from the mSUGRA gaugino mass relations are within 5\%.
However, in ${\cal F}$-$SU(5)$, we predict the deviations from the mSUGRA gaugino mass relations to be
larger due to the presence of vector-like particles, which can be tested at the colliders.
We determine the viable parameter space that satisfies all the latest experimental
constraints and find it is consistent with the CDMS II experiment. Further, we compute the
cross-sections of neutralino annihilations into gamma-rays and compare to the first published
Fermi-LAT measurement. Finally, the corresponding range of proton lifetime predictions is calculated and found
to be within reach of the future Hyper-Kamiokande and DUSEL experiments.

\end{abstract}


\preprint{ACT-04-10, MIFPA-10-13}

\maketitle

\section{Introduction}


As we initiate the era of the Large Hadron Collider (LHC), we await with anticipation the expected discovery of supersymmetry
and the Higgs states required to break electroweak symmetry and stabilize the electroweak scale.
On the other hand, there is thus far no concrete model that can explain all observed physics in a comprehensive mathematical framework.
Unique predictions that can be tested at the LHC, future International Linear Collider (ILC),
and other forthcoming experiments are necessary if string theory is to be substantiated as the correct fundamental description of nature.
Following a top-down approach, it may be feasible to derive all known observable physics from a fundamental theory such as string theory.
In contrast, the bottom-up approach offers the possibility to infer the framework of the fundamental theory at high-energy from
a low-energy signal at the experiments. In the spirit of this bottom-up approach, our goal here is to study Grand Unified Theories (GUTs)
from F-Theory, which have seen exciting progress the past two years, and present F-Theory GUT low-energy physics observable at current and future experiments.

F-Theory can be considered as the
strongly coupled formulation of ten-dimensional Type IIB string 
theory with a varying axion ($a$)-dilaton ($\phi$) field 
$S=a+ie^{-\phi}$~\cite{Vafa:1996xn}. GUTs were first locally constructed 
in F-theory two years ago~\cite{Donagi:2008ca, Beasley:2008dc, Beasley:2008kw,
Donagi:2008kj}, and subsequently, model
building and phenomenological consequences have been studied 
extensively~\cite{Heckman:2008es, Heckman:2008qt, Font:2008id,
Heckman:2008qa, Jiang:2009zza, Blumenhagen:2008aw, Heckman:2008jy, Hayashi:2009ge,
Chen:2009me, Heckman:2009bi, Donagi:2009ra, Jiang:2009za, Li:2009cy, 
Marsano:2009gv, Cecotti:2009zf, Li:2009fq, Li:2010dp, Marsano:2009wr, 
Leontaris:2009wi, Heckman:2010xz, Li:2010mr}. 
In F-theory model building, 
the gauge fields are on the observable seven-branes that wrap a del Pezzo 
$n$ ($dP_n$) surface for the extra four space dimensions, 
the SM fermions and Higgs fields are localized on the complex
codimension-one curves (two-dimensional real subspaces) in $dP_n$,
and the Standard Model (SM) fermion Yukawa couplings arise 
from the triple intersections of the
SM fermion and Higgs curves.
A brand new feature is that the $SU(5)$ gauge symmetry
can be broken down to the SM gauge symmetry
by turning on the $U(1)_Y$ 
flux~\cite{Beasley:2008dc, Beasley:2008kw, Li:2009cy}, 
and the $SO(10)$ gauge 
symmetry can be broken down to the $SU(5)\times U(1)_X$
and $SU(3)\times SU(2)_L\times SU(2)_R\times U(1)_{B-L}$
gauge symmetries by turning on the $U(1)_X$ and $U(1)_{B-L}$
fluxes, respectively~\cite{Beasley:2008dc, Beasley:2008kw,
Jiang:2009zza, Jiang:2009za, Font:2008id, Li:2009cy}. 

Intriguingly, the $SO(10)$ models have both gauge interaction
unification and SM fermion unification, and consequently we believe
the $SO(10)$ models are more interesting to study than the 
$SU(5)$ models. Moreover, we can break the
$SO(10)$ gauge symmetry down to the flipped $SU(5)\times U(1)_X$
gauge symmetry by turning on the $U(1)_X$ flux, wherein 
the doublet-triplet splitting problem can be solved
elegantly~\cite{smbarr, dimitri, AEHN-0}. 
The flipped $SU(5)\times U(1)_X$ models with TeV scale 
vector-like particles~\cite{Jiang:2006hf},
which have been dubbed ``${\cal F}$-$SU(5)$'',
have been constructed systematically in F-theory~\cite{Jiang:2009zza, Jiang:2009za}. 
Models of this nature can be also obtained in the free fermionic string constructions~\cite{Lopez:1992kg}. In ${\cal F}$-$SU(5)$,
the $SU(3)_C\times SU(2)_L$ gauge symmetries are unified at
about $10^{16}$ GeV, and the $SU(5)\times U(1)_X$ gauge symmetries
are unified above $10^{17}$ GeV. Thus, we can solve the
monopole problem. On top of this, the TeV scale vector-like particles
are potentially observable at the Large Hadron Collider (LHC), 
the lightest CP-even Higgs boson mass
can be lifted~\cite{HLNT-P}, and the predicted proton decay~\cite{Ellis:1995at,Ellis:2002vk,Li:2009fq,Li:2010dp}
is within the reach of the future 
Hyper-Kamiokande~\cite{Nakamura:2003hk} and
Deep Underground Science and Engineering 
Laboratory (DUSEL)~\cite{DUSEL} experiments.

Recently, the Cryogenic Dark Matter Search (CDMS) collaboration
observed two candidate dark matter events 
in the CDMS II experiment~\cite{Ahmed:2009zw}. 
The recoil energies for these two events were 12.3 keV 
and 15.5 keV, respectively, and the data set 
an upper limit on the dark matter-nucleon elastic-scattering spin
independent cross section around $10^{-8}-10^{-7}$ pb.
The probability of observing two or more background events is
$23\%$, so the CDMS II results cannot be considered statistically
significant evidence for dark matter interactions, although
there is some strong possibility that they do in fact represent an 
authentic signal. In particular, the favored
dark matter mass from the CDMS II data is about 100 GeV. 
Interestingly, the CDMS II experiment can be explained in 
the supersymmetric Standard Model with $R$ parity
where the lightest supersymmetric particle (LSP) neutralino is 
the dark matter candidate~\cite{Bernal:2009jc, Bottino:2009km, 
Feldman:2009pn, Allahverdi:2009sb, Endo:2009uj, Holmes:2009uu, Hisano:2009xv,
Gogoladze:2009mc, Maxin:2009kp, Gao:2010pg}.

In this paper, we first briefly review ${\cal F}$-$SU(5)$, and then subsequently
study the supersymmetry breaking soft terms in
the framework of gravity mediation. Although the gaugino masses are not unified
at the traditional GUT scale, we still have the same gaugino mass relation
at the electroweak scale as the minimal supergravity (mSUGRA) scenario at one 
loop~\cite{mSUGRA}. We also introduce two parameters testable at the colliders 
to measure the small 
two-loop deviations from the mSUGRA gaugino mass relations at the electroweak scale. 
Next, we incorporate TeV scale vector particles and generate regions of the parameter space that can satisfy all current experimental
constraints and are consistent with the CDMS II experiment~\cite{Ahmed:2009zw}. 
In addition, we compute the annihilation cross-section of two neutralinos into two gamma-rays and evaluate the results in light of the first published Fermi-LAT measurement. Then we compute the new parameters to measure the two-loop deviations for both ${\cal F}$-$SU(5)$ and mSUGRA. 
We find that the deviations from the mSUGRA gaugino mass 
relation in mSUGRA are smaller than 5\%, while the deviations in the ${\cal F}$-$SU(5)$ 
are larger as a result of the TeV scale vector-like particles.
An analytical comparison of the deviation between mSUGRA and ${\cal F}$-$SU(5)$ is illustrated. Next, the expected observable final states at the LHC are given for all viable regions of the parameter space. Lastly, the proton lifetime is calculated for the experimentally allowed regions of the ${\cal F}$-$SU(5)$ parameter space, and we show that the rate of predicted decay is indeed within the reach of the future Hyper-Kamiokande and DUSEL experiments.


\section{${\cal F}$-$SU(5)$}

First, we
briefly review the minimal flipped
$SU(5)\times U(1)_X$ model~\cite{smbarr, dimitri, AEHN-0}. 
The gauge group of the flipped $SU(5)$ model is
$SU(5)\times U(1)_{X}$, which can be embedded into $SO(10)$.
We define the generator $U(1)_{Y'}$ in $SU(5)$ as 
\bea 
T_{\rm U(1)_{Y'}}={\rm diag} \left(-\frac{1}{3}, -\frac{1}{3}, -\frac{1}{3},
 \frac{1}{2},  \frac{1}{2} \right).
\label{u1yp}
\eea
The hypercharge is given by
\bea
Q_{Y} = \frac{1}{5} \left( Q_{X}-Q_{Y'} \right).
\label{ycharge}
\eea

There are three families of SM fermions 
whose quantum numbers under $SU(5)\times U(1)_{X}$ are
\bea
F_i={\mathbf{(10, 1)}},~ {\bar f}_i={\mathbf{(\bar 5, -3)}},~
{\bar l}_i={\mathbf{(1, 5)}},
\label{smfermions}
\eea
where $i=1, 2, 3$. 

To break the GUT and electroweak gauge symmetries, we 
introduce two pairs of Higgs fields
\bea
H={\mathbf{(10, 1)}},~{\overline{H}}={\mathbf{({\overline{10}}, -1)}},
~h={\mathbf{(5, -2)}},~{\overline h}={\mathbf{({\bar {5}}, 2)}},
\label{Higgse1}
\eea
where particle assignments of the Higgs fields are 
\bea
H=(Q_H, D_H^c, N_H^c)~,~
{\overline{H}}= ({\overline{Q}}_{\overline{H}}, {\overline{D}}^c_{\overline{H}}, 
{\overline {N}}^c_{\overline H})~,~\,
\label{Higgse2} \\
h=(D_h, D_h, D_h, H_d)~,~
{\overline h}=({\overline {D}}_{\overline h}, {\overline {D}}_{\overline h},
{\overline {D}}_{\overline h}, H_u)~,~\,
\label{Higgse3}
\eea
and $H_d$ and $H_u$ are one pair of Higgs doublets in the supersymmetric SM.
We also add a SM singlet field $\Phi$.

To break the $SU(5)\times U(1)_{X}$ gauge symmetry,
we introduce the following Higgs superpotential 
\bea
{\it W}~=~\lambda_1 H H h + \lambda_2 {\overline H} {\overline H} {\overline
h} + \Phi ({\overline H} H-M_{\rm H}^2). 
\label{spgut} 
\eea 
There is only
one F-flat and D-flat direction, which can always be rotated
such that $\langle N^c_H\rangle = \langle{\overline {N}}^c_{\overline H}\rangle = M_{\rm H}$.
In addition, the superfields $H$ and ${\overline H}$ are absorbed, acquiring large masses via
the supersymmetric Higgs mechanism, except for $D_H^c$ and 
${\overline {D}}^c_{\overline H}$. The superpotential terms 
$ \lambda_1 H H h$ and
$ \lambda_2 {\overline H} {\overline H} {\overline h}$ couple the $D_H^c$ and
${\overline {D}}^c_{\overline H}$ with the $D_h$ and ${\overline {D}}_{\overline h}$,
respectively, to form heavy eigenstates with masses
$2 \lambda_1 \langle N_H^c\rangle$ and $2 \lambda_2 \langle{\overline {N}}^c_{\overline H}\rangle$. In consequence, we
naturally achieve doublet-triplet splitting due to the missing
partner mechanism~\cite{AEHN-0}. The triplets in $h$ and ${\overline h}$ only have
small mixing through the $\mu h {\overline h}$ term with
$\mu$ around the TeV scale, so we also solve the dimension five
proton decay problem from the colored Higgsino exchange.

In flipped $SU(5)\times U(1)_X$ models, the 
$SU(3)_C\times SU(2)_L$ gauge couplings are first joined 
at the scale $M_{32}$, and the $SU(5)$ and $U(1)_X$ gauge
couplings are subsequently unified at the higher scale $M_U$. To separate the $M_{32}$ and $M_U$ scales
and obtain true string-scale gauge coupling unification in 
free fermionic string models~\cite{Jiang:2006hf} or
the decoupling scenario in F-theory models~\cite{Jiang:2009zza, Jiang:2009za},
we introduce vector-like particles which form complete
flipped $SU(5)\times U(1)_X$ multiplets.
In order to avoid the Landau pole
problem for the strong coupling constant, we can only introduce the
following two sets of vector-like particles around the TeV 
scale~\cite{Jiang:2006hf}
\begin{eqnarray}
&& Z1:  XF ={\mathbf{(10, 1)}}~,~
{\overline{XF}}={\mathbf{({\overline{10}}, -1)}}~;~\\
&& Z2: XF~,~{\overline{XF}}~,~Xl={\mathbf{(1, -5)}}~,~
{\overline{Xl}}={\mathbf{(1, 5)}}
~.~\,
\end{eqnarray}
For notational simplicity, we define the flipped
$SU(5)\times U(1)_X$ models with $Z1$ and
$Z2$ sets of vector-like particles as 
Type I and Type II flipped
$SU(5)\times U(1)_X$ models, respectively.
We choose to focus in this paper on the Type II model,
although results are similar in most regards for Type I.
We emphasize that the Type I and II flipped $SU(5)\times U(1)_X$
models have been constructed consistently from F-theory in
Refs.~\cite{Jiang:2009zza, Jiang:2009za}.

Next, let us consider gravity mediated supersymmetry breaking.
With the $U(1)_X$ flux, the gauge kinetic functions $f_5$ and $f_{1X}$, 
respectively for $SU(5)$ and $U(1)_X$ gauge symmetry, are~\cite{Jiang:2009za}
\begin{eqnarray}
f_5 ~=~ f_{1X}~=~ \tau + a S~,~\,
\end{eqnarray}
where $a$ is an integer.
The first term is the original gauge kinetic function for
$SO(10)$, and the second term arises from the $U(1)_X$ flux
effects. After the $SU(5)\times U(1)_X$ gauge symmetries
are broken down to the SM gauge symmetry, the gauge kinetic function $f_1$ for $U(1)_Y$ is
\begin{eqnarray}
f_1 ~\equiv~ {\frac{24}{25}} f_{1X} + {\frac{1}{25}}f_5 ~.~\,
\label{U(1)Y-GKF}
\end{eqnarray}

In the gravity mediated supersymmetry breaking scenario where
the F-terms of $\tau$ and $S$ break supersymmetry, 
we obtain the gaugino mass relation from the
scale $M_U$ down to the scale $M_{32}$~\cite{Li:2010xr}
\begin{eqnarray}
{\frac{M_5}{\alpha_5}}~=~{\frac{M_{1X}}{\alpha_{1X}}}~,~\,
\label{M51-GMR}
\end{eqnarray}
where $M_5$ and $M_{1X}$ are gaugino masses for
$SU(5)$ and $U(1)_X$ respectively, and $\alpha_5$ and $\alpha_{1X}$ are
the $SU(5)$ and $U(1)_X$ gauge couplings.
This gaugino mass relation is renormalization scale invariant
under one-loop renormalization group equation (RGE) running.

Using Eq.~(\ref{U(1)Y-GKF}), we obtain the gaugino mass $M_1$ for $U(1)_Y$ at the scale $M_{32}$
\begin{eqnarray}
 {\frac{M_1}{\alpha_1}} &\equiv & {\frac{24}{25}} {\frac{M_{1X}}{\alpha_{1X}}}
+ {\frac{1}{25}} {\frac{M_5}{\alpha_5}} ~,~\,
\end{eqnarray}
where $\alpha_1\equiv 5\alpha_{Y}/3$ is the $U(1)_Y$ gauge coupling.
With Eq.~(\ref{M51-GMR}), we obtain at the scale $M_{32}$
\begin{eqnarray}
 {\frac{M_1}{\alpha_1}} &=& {\frac{M_5}{\alpha_5}}~.~\,
\label{GauginoMR}
\end{eqnarray}
Note that the gauge couplings $\alpha_1$ and $\alpha_5$ are split,
and thus we do not have universal gaugino masses at the usual GUT scale.
Applying RGE running, we obtain the gaugino mass relation
which is valid from the scale $M_{32}$ down to the electroweak scale
at one loop~\cite{Li:2010xr}
\begin{eqnarray}
{\frac{M_3}{\alpha_3}} ~=~  {\frac{M_2}{\alpha_2}} 
~=~  {\frac{M_1}{\alpha_1}} ~,~\,
\label{mSUGRA}
\end{eqnarray}
where $\alpha_3$ and $\alpha_2$ are
couplings of the $SU(3)_C$ and $SU(2)_L$ gauge symmetries respectively, and 
$M_3$ and $M_2$ are masses of the $SU(3)_C$ and $SU(2)_L$ 
gauginos. This gaugino mass relation is the same as
that in mSUGRA~\cite{mSUGRA}. The gaugino masses can be
measured at the LHC and future ILC~\cite{Cho:2007qv, Barger:1999tn}, 
hence, we can test this relation at future
experiments. 

Considering two-loop RGE running, we will have a small deviation for
the gaugino mass relation given by Eq.~(\ref{mSUGRA}). To quantify the
deviations, we first define
\begin{eqnarray}
\left({\frac{M_i}{\alpha_i}}\right)_{\rm L} & \equiv & {\rm maximum}
\left[{\frac{M_3}{\alpha_3}},~ {\frac{M_2}{\alpha_2}},
~{\frac{M_1}{\alpha_1}}\right] ~,~\, \nonumber \\
\left({\frac{M_i}{\alpha_i}}\right)_{\rm M} & \equiv & {\rm median}
\left[{\frac{M_3}{\alpha_3}},~ {\frac{M_2}{\alpha_2}},
~{\frac{M_1}{\alpha_1}}\right] ~,~\, \nonumber \\
\left({\frac{M_i}{\alpha_i}}\right)_{\rm S} & \equiv & {\rm minimum}
\left[{\frac{M_3}{\alpha_3}},~ {\frac{M_2}{\alpha_2}},
~{\frac{M_1}{\alpha_1}}\right] ~.~\,
\end{eqnarray}
Then, we define the small deviations as follows
\begin{eqnarray}
\delta_{+} ~=~ {\frac{\left({\frac{M_i}{\alpha_i}}\right)_{\rm L}
-\left({\frac{M_i}{\alpha_i}}\right)_{\rm M}}
{\left({\frac{M_i}{\alpha_i}}\right)_{\rm M}}}~,~~~
\delta_{-} ~=~ {\frac{\left({\frac{M_i}{\alpha_i}}\right)_{\rm S}
-\left({\frac{M_i}{\alpha_i}}\right)_{\rm M}}
{\left({\frac{M_i}{\alpha_i}}\right)_{\rm M}}}
~.~\,
\label{delta}
\end{eqnarray}
We emphasize that $\delta_+ \ge 0$ and $\delta_+ \le 0$.
In other words, we have
\begin{eqnarray}
\left({\frac{M_i}{\alpha_i}}\right)_{\rm L} ~:~
\left({\frac{M_i}{\alpha_i}}\right)_{\rm M} ~:~
\left({\frac{M_i}{\alpha_i}}\right)_{\rm S} &=&
(1+\delta_+)  ~:~ 1  ~:~ (1+\delta_-)
~.~\,
\end{eqnarray}

For example, in $\cal{F}$-$SU(5)$ we have
\begin{eqnarray}
\left({\frac{M_i}{\alpha_i}}\right)_{\rm L} ~=~{\frac{M_3}{\alpha_3}}
~,~~~
\left({\frac{M_i}{\alpha_i}}\right)_{\rm M} ~=~{\frac{M_2}{\alpha_2}}
~,~~~
\left({\frac{M_i}{\alpha_i}}\right)_{\rm S} ~=~{\frac{M_1}{\alpha_1}}
~.~\,
\label{FSUV-GM}
\end{eqnarray}

\section{Low Energy Supersymmetry Phenomenology}

The ${\cal F}$-$SU(5)$ type models have been constructed locally in F-theory~\cite{Jiang:2009zza, Jiang:2009za}, and thus we do not know the K\"ahler potential for the SM fermions and Higgs fields and cannot calculate the supersymmetry breaking scalar masses and trilinear soft terms. Interestingly, as long as the scalar masses and trilinear soft terms are around the TeV scale, the gaugino mass relation can be preserved very well at low energy to two loops. For simplicity, in this paper, we consider the universal scalar mass $m_0$ and the universal trilinear soft term $A_{0}$.
Essentially speaking, we study the F-theory inspired low energy supersymmetry phenomenology, so that we have four free parameters in our model: 
$M_{5}$, $m_0$, $A_{0}$, and $\tan\beta$, where $\tan\beta=\langle H_u \rangle / \langle H_d \rangle$, the ratio of the vacuum expectation values of the Higgs scalar fields. In addition, we must choose the sign of $\mu$. We take $\mu > 0$, as suggested by the results of $g_{\mu}-2$ for the muon. We also assume that the masses for the vector-like particles are universal, at 1 TeV. The contributions to the beta functions of the SM gauge couplings from the vector-like particles must be accounted for in the one-loop and two-loop gauge coupling RGEs when running the soft terms down to the electroweak scale. We alter the one-loop and two-loop gauge coupling RGE code in {\tt SuSpect 2.34}~\cite{Djouadi:2002ze} to realize the beta 
function corrections. In an effort to minimize the complexity of revising the code, we use only the one-loop RGEs for the SM fermion Yukawa couplings and supersymmetry breaking soft terms. The Yukawa coupling RGEs contribute to the gauge coupling RGEs at second order, so it is consistent to run only the Yukawa coupling RGEs at one-loop. However, for future investigations of GUTs with vector-like particles, we intend to extend our analysis to include two-loop RGEs for all supersymmetry breaking terms. Next, we modify the one-loop gaugino mass RGEs to include the strong coupling effects of the vector-like particles. Finally, the {\tt SuSpect} code must be revised to require gauge coupling unification at $g_{2} = g_{3}$, rather than the default unification configuration of $g_{1} = g_{2}$. This necessitates lowering the ceiling on the running of the gauge couplings at high scale due to large effects on $g_{2}$ at scales above $10^{18}$ GeV. We see in Fig.~\ref{fig:running_plot} the split in unification at the $M_{32}$ scale, as explained in the previous section. Below 1 TeV, there is only the running of the SM gauge couplings, which are shown as the red dotted lines in Fig.~\ref{fig:running_plot} for an unflipped $SU(5)$ point with universal gaugino masses, however, at 1 TeV and higher energies, we see the effects of the vector-like particles on the running of the couplings and masses. The relic LSP neutralino density and WIMP-nucleon direct detection cross-sections are computed with {\tt MicrOMEGAs 2.1}~\cite{Belanger:2008sj,Belanger:2006is}, although the {\tt MicrOMEGAs} code must also be reworked to include the revised RGEs. We use {\tt SuSpect} as the RGE code in our relic density and WIMP-nucleon cross-section computations, and hence we implement the revised {\tt SuSpect} RGEs in {\tt MicrOMEGAs} as well. Supersymmetry breaking soft terms for the present model are generated using the gaugino mass relation in Eq.~(\ref{GauginoMR}) at the scale $M_{32}$. The gaugino mass $M_{1}$ is dependent upon the ratio of the unification scale gauge couplings $\alpha_{1}$ and $\alpha_{5}$, therefore, an iterative procedure must first be used to determine the final value of $M_{1}$ before any low energy phenomenology can be investigated. We use a top quark mass of $m_{t}$ = 173.1 GeV~\cite{:2009ec} and employ the following experimental constraints:

\begin{figure}[ht]
	\centering
		\includegraphics[width=0.75\textwidth]{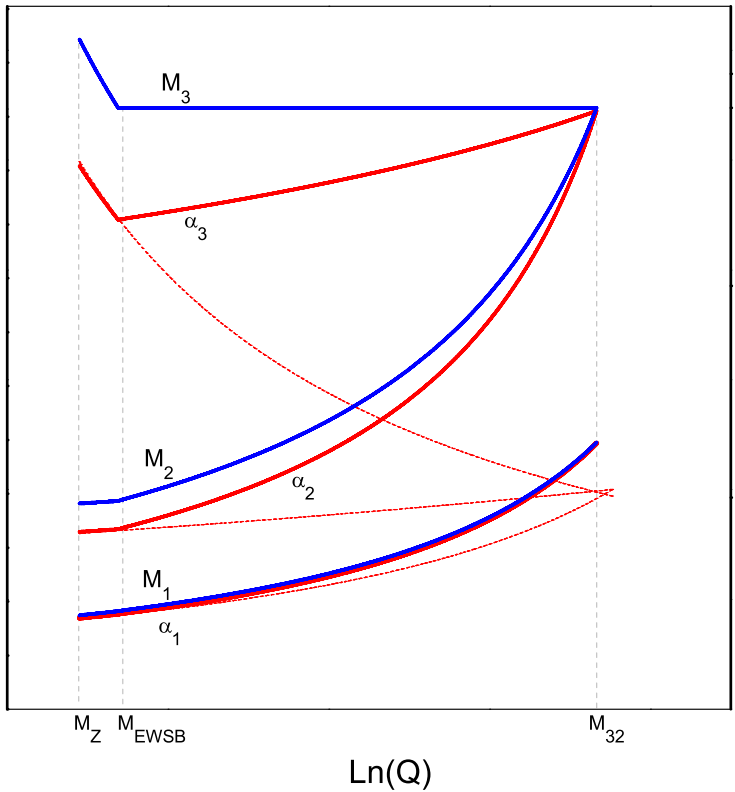}
		\caption{Running of the ${\cal F}$-$SU(5)$ gaugino masses and gauge couplings from the $M_{Z}$ scale to the $M_{32}$ partial unification scale. The red dotted line represents the running of the gauge couplings for unflipped $SU(5)$ with universal gaugino masses at $M_{GUT}$. We suppress labeling of the vertical axes to preserve general heuristic applicability to the full parameter space considered.}
	\label{fig:running_plot}
\end{figure}

\begin{enumerate}

\item The WMAP 2$\sigma$ measurements of the cold dark matter density~\cite{Spergel:2006hy}, 0.095 $\leq \Omega_{\chi} \leq$ 0.129. In addition, we look at the Supercritical String  Cosmology (SSC) model~\cite{Antoniadis:1988aa}, in which a dilution factor of $\cal{O}$(10) is allowed~\cite{Lahanas:2006hf}, where $\Omega_{\chi} \lesssim$ 1.1. For a discussion of the SSC model within the context of mSUGRA, see~\cite{Dutta:2008ge}. We also investigate another case where a neutralino LSP makes up a subdominant component and employ this possibility by removing the lower bound.

\item The experimental limits on the Flavor Changing Neutral Current (FCNC) process, $b \rightarrow s\gamma$. The results from the Heavy Flavor Averaging Group (HFAG)~\cite{Barberio:2007cr}, in addition to the BABAR, Belle, and CLEO results, are: $Br(b \rightarrow s\gamma) = (355 \pm 24^{+9}_{-10} \pm 3) \times 10^{-6}$. There is also a more recent estimate~\cite{Misiak:2006zs} of $Br(b \rightarrow s\gamma) = (3.15 \pm 0.23) \times 10^{-4}$. For our analysis, we use the limits $2.86 \times 10^{-4} \leq Br(b \rightarrow s\gamma) \leq 4.18 \times 10^{-4}$, where experimental and
theoretical errors are added in quadrature.

\item The anomalous magnetic moment of the muon, $g_{\mu} - 2$. For this analysis we use the 2$\sigma$ level boundaries, $11 \times 10^{-10} < a_{\mu} < 44 \times 10^{-10}$~\cite{Bennett:2004pv}.

\item The process $B_{s}^{0} \rightarrow \mu^+ \mu^-$ where the decay has a ${\tan}^6\beta$ dependence. We take the upper bound to be $Br(B_{s}^{0} \rightarrow \mu^{+}\mu^{-}) < 5.8 \times 10^{-8}$~\cite{:2007kv}.

\item The LEP limit on the lightest CP-even Higgs boson mass, $m_{h} \geq 114$ GeV~\cite{Barate:2003sz}.

\end{enumerate}

\begin{figure}[ht]
	\centering
		\includegraphics[width=0.75\textwidth]{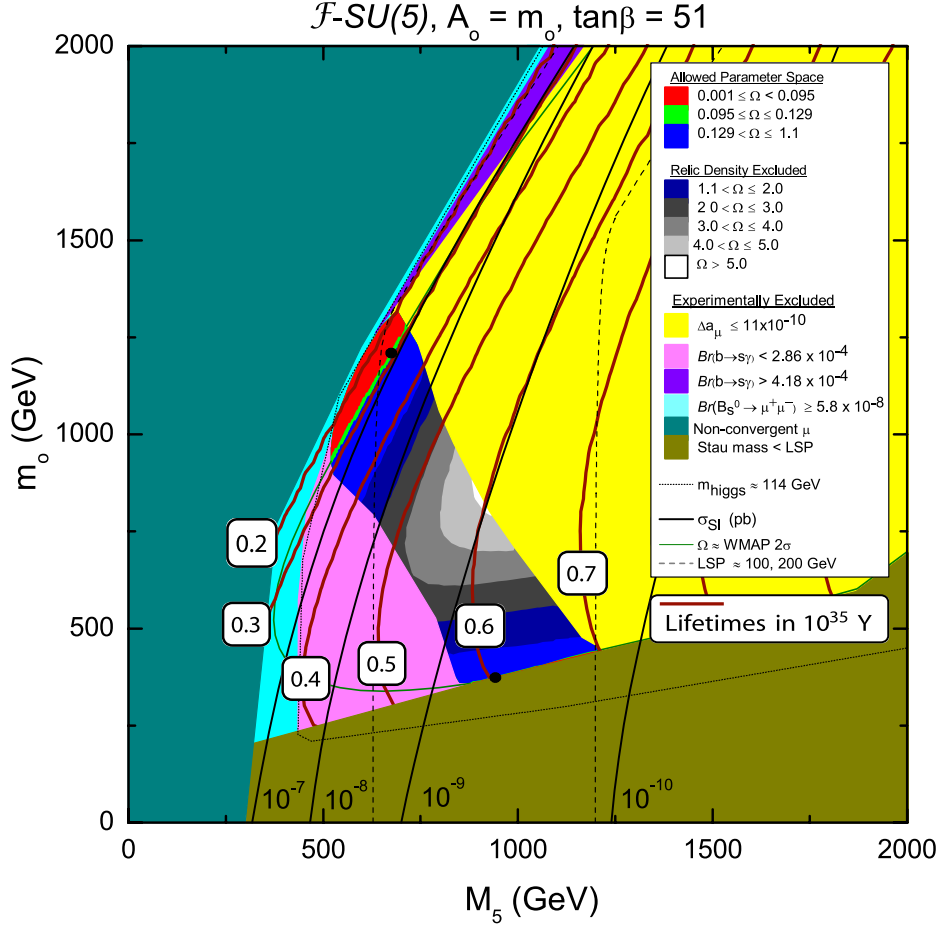}
		\caption{${\cal F}$-$SU(5)$ experimentally allowed parameter space for $\tan \beta = 51$, $A_{0} = m_{0}$. The benchmark points detailed in Table~\ref{tab:masses1} and Table~\ref{tab:masses2} are annotated by the black points. The solid black lines demarcate the WIMP-nucleon spin-independent cross-sections. The dotted line indicates the LEP boundary for the light Higgs. The two dashed lines represent an LSP mass of 100 GeV and 200 GeV. The dark green line denotes the region satisfying the WMAP 2$\sigma$ relic density, however, only the bright green region within the designated allowed parameter space can fulfill all the experimental constraints.  The dark red contours label $p \!\rightarrow\! {(e\vert\mu)}^{\!+}\! \pi^0$ proton lifetime predictions, in units of $10^{35}$ years.}
	\label{fig:movsm5_plot}
\end{figure}

We commence with $M_{5}$, $m_{0}$, $A_{0}$, and $\tan \beta$ as free parameters, however, for simplicity we take $A_{0} = m_{0}$. A comprehensive scan of the entire parameter space uncovers $\tan \beta = 7$ to $\tan \beta = 53$ as consistent with all the experimental constraints and latest CDMS II results~\cite{Ahmed:2009zw}, thus we choose tan$\beta$ = 51 for our benchmark point to study in this work. As shown in Fig.~\ref{fig:movsm5_plot}, the experimentally allowed parameter space for $\tan \beta = 51$ after applying all these constraints consists of two segregated small regions. We look to examine more closely those regions of the experimentally allowed parameter space that comply with the recent CDMS II data and the WMAP relic density observations. This encompasses the narrow WMAP strip in both experimentally allowed regions, so we choose points from each allowed region within these narrow WMAP strips as our benchmark points for analysis. In fact, with relic densities of $\Omega_{\chi}$ = 0.1093 and $\Omega_{\chi}$ = 0.1151, these benchmark points additionally satisfy the very constrained WMAP 7-year~\cite{Komatsu:2010fb} results. Inspecting the sparticle and Higgs spectrum for our benchmark point in Table~\ref{tab:masses1} with relic density $\Omega_{\chi}$ = 0.1093 reveals that the additional contribution of the 1 TeV vector-like particles lower the gluino mass quite dramatically. The gaugino mass $M_{3}$ runs flat from the $M_{32}$ unification scale to the electroweak scale as shown in Fig.~\ref{fig:running_plot}, though, due to SUSY radiative corrections, the physical gluino mass is larger than $M_{3}$ at the $M_{32}$ scale. This is true for all regions of the ${\cal F}$-$SU(5)$ parameter space. In mSUGRA, the focus point region consists of large $m_{0}$ where the WMAP observed relic density can be satisfied with a mixed Bino-Higgsino state in the LSP due to a small $|\mu|$, leading to enhanced $\widetilde{\chi}_{1}^{0} \widetilde{\chi}_{1}^{0}$ annihilation. However, even though $m_{0}$ is reasonably large for the Table~\ref{tab:masses1} benchmark point, here the LSP is 99\% Bino for ${\cal F}$-$SU(5)$. In contrast, the sparticle and Higgs spectrum for our benchmark point in Table~\ref{tab:masses2} illustrates that the WMAP relic density $\Omega_{\chi} = 0.1151$ is generated through stau-neutralino coannihilation. This is demonstrated by the near degenerate mass between the $\widetilde{\chi}_{1}^{0}$ neutralino LSP and $\widetilde{\tau}_{1}$ stau next-to-lightest supersymmetric particle (NLSP). Accordingly, the LSP for the benchmark point in Table~\ref{tab:masses2} is 99.9\% bino.

\begin{table}[ht]
	\centering
	\caption{Sparticle and Higgs spectrum for the $M_{5}$ = 670 GeV and $m_{0}$ = 1215 GeV benchmark point illustrated in Fig.~\ref{fig:movsm5_plot}. Here, $\tan \beta = 51$, $\Omega_{\chi}$ = 0.1093, $\sigma_{SI} = 1.4 \times 10^{-7}$ pb, and $\left\langle \sigma v \right\rangle_{\gamma\gamma} = 2.0 \times 10^{-26} ~cm^{3}/s$. The GUT-scale mass parameters for this point are (in GeV) $M_{3}$ = 670, $M_{2}$ = 670, $M_{1}$ = 296, $A_{0} = m_{0}$ = 1215.  The central prediction for the $p \!\rightarrow\! {(e\vert\mu)}^{\!+}\! \pi^0$ proton lifetime is $2.8\times 10^{34}$ years.}
		\begin{tabular}{|c|c||c|c|} \hline
    Sparticle & Mass~(GeV) & Sparticle & Mass~(GeV)\\ \hline\hline		
    $\widetilde{\chi}_{1}^{0}$&$107$&$\widetilde{t}_{1}$&$1206$\\ \hline
    $\widetilde{\chi}_{2}^{0}$&$224$&$\widetilde{t}_{2}$&$1359$\\ \hline
    $\widetilde{\chi}_{3}^{0}$&$337$&$\widetilde{b}_{1}$&$1341$\\ \hline
    $\widetilde{\chi}_{4}^{0}$&$363$&$\widetilde{b}_{2}$&$1374$\\ \hline
    $\widetilde{\chi}_{1}^{\pm}$&$223$&$\widetilde{u}_{R}$&$1627$\\ \hline
    $\widetilde{\chi}_{2}^{\pm}$&$363$&$\widetilde{u}_{L}$&$1670$\\ \hline
    $\widetilde{\tau}_{1}$&$822$&$\widetilde{d}_{R}$&$1626$\\ \hline
    $\widetilde{\tau}_{2}$&$1107$&$\widetilde{d}_{L}$&$1672$\\ \hline
    $\widetilde{e}_{R}$&$1221$&$\widetilde{g}$&$817$\\ \hline
    $\widetilde{e}_{L}$&$1276$&$m_{h}$&$116.6$\\ \hline
    $\widetilde{\nu}_{e/ \mu}$&$1274$&$m_{H,A}$&$337$\\ \hline
    $\widetilde{\nu}_{\tau}$&$1103$&$m_{H^{\pm}}$&$349$\\ \hline
		\end{tabular}
		\label{tab:masses1}
\end{table}

\begin{table}[ht]
	\centering
	\caption{Sparticle and Higgs spectrum for the $M_{5}$ = 925 GeV and $m_{0}$ = 375 GeV benchmark point illustrated in Fig.~\ref{fig:movsm5_plot}. Here, $\tan \beta = 51$, $\Omega_{\chi} = 0.1151$, $\sigma_{SI} = 5.4 \times 10^{-10}$ pb, and $\left\langle \sigma v \right\rangle_{\gamma\gamma} = 1.1 \times 10^{-27} ~cm^{3}/s$. The GUT-scale mass parameters for this point are (in GeV) $M_{3}$ = 925, $M_{2}$ = 925, $M_{1}$ = 418, $A_{0} = m_{0}$ = 375.  The central prediction for the $p \!\rightarrow\! {(e\vert\mu)}^{\!+}\! \pi^0$ proton lifetime is $6.0\times 10^{34}$ years.}
		\begin{tabular}{|c|c||c|c|} \hline
    Sparticle & Mass~(GeV) & Sparticle & Mass~(GeV)\\ \hline\hline		
    $\widetilde{\chi}_{1}^{0}$&$152$&$\widetilde{t}_{1}$&$1122$\\ \hline
    $\widetilde{\chi}_{2}^{0}$&$336$&$\widetilde{t}_{2}$&$1375$\\ \hline
    $\widetilde{\chi}_{3}^{0}$&$1027$&$\widetilde{b}_{1}$&$1289$\\ \hline
    $\widetilde{\chi}_{4}^{0}$&$1030$&$\widetilde{b}_{2}$&$1378$\\ \hline
    $\widetilde{\chi}_{1}^{\pm}$&$336$&$\widetilde{u}_{R}$&$1523$\\ \hline
    $\widetilde{\chi}_{2}^{\pm}$&$1031$&$\widetilde{u}_{L}$&$1609$\\ \hline
    $\widetilde{\tau}_{1}$&$168$&$\widetilde{d}_{R}$&$1521$\\ \hline
    $\widetilde{\tau}_{2}$&$634$&$\widetilde{d}_{L}$&$1611$\\ \hline
    $\widetilde{e}_{R}$&$408$&$\widetilde{g}$&$1060$\\ \hline
    $\widetilde{e}_{L}$&$656$&$m_{h}$&$118.3$\\ \hline
    $\widetilde{\nu}_{e/ \mu}$&$651$&$m_{H,A}$&$732$\\ \hline
    $\widetilde{\nu}_{\tau}$&$607$&$m_{H^{\pm}}$&$737$\\ \hline
		\end{tabular}
		\label{tab:masses2}
\end{table}

\begin{figure}[ht]
	\centering
		\includegraphics[width=0.75\textwidth]{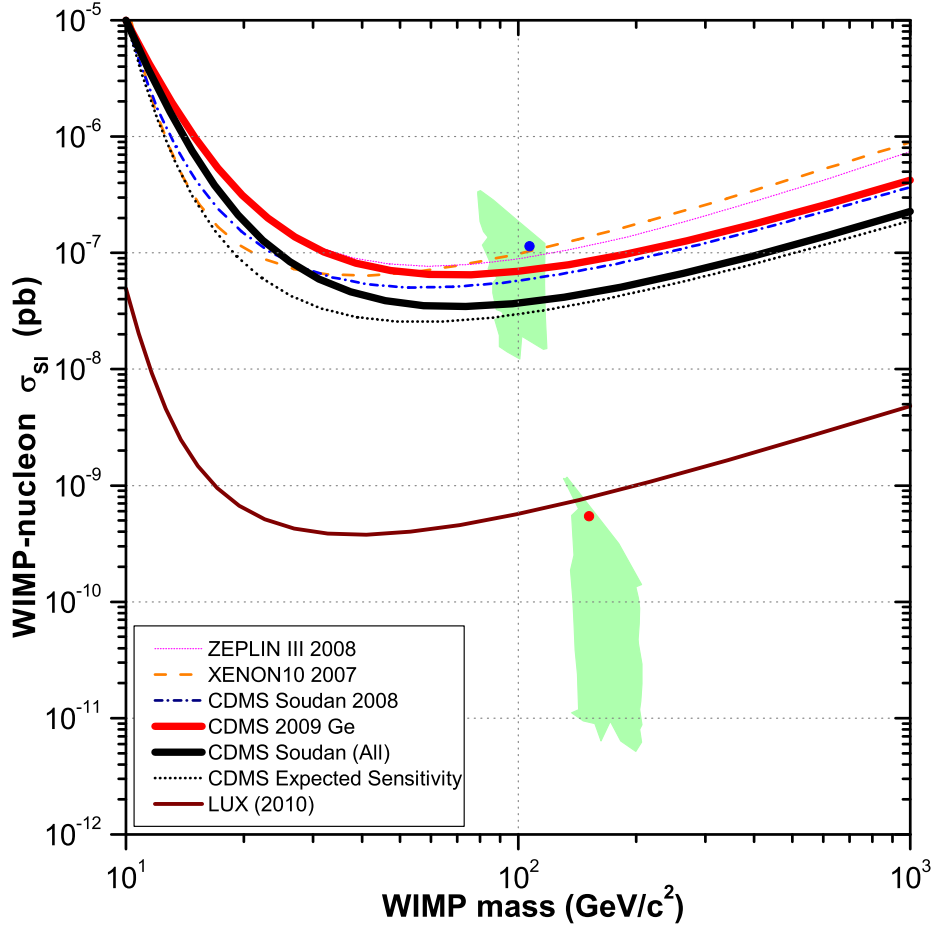}
		\caption{${\cal F}$-$SU(5)$ spin-independent WIMP-nucleon cross-sections for $\tan \beta = 51$, $A_{0} = m_{0}$, overlaid with direct-detection limits from recent and forthcoming experiments. The shaded regions satisfy all experimental constraints. The benchmark point in Table~\ref{tab:masses1} is annotated by the blue point, while the benchmark point in Table~\ref{tab:masses2} is annotated by the red point.}
	\label{fig:sigma_plot}
\end{figure}

The WIMP-nucleon direct-detection cross-sections $\sigma_{SI}$ depicted in Fig.~\ref{fig:sigma_plot} underscore the fact that the case of $\tan \beta = 51$ produces WIMPs with $\sigma_{SI}$ that comply with the CDMS II upper limits, with our benchmark point in Table~\ref{tab:masses1} at $\sigma_{SI} = 1.4 \times 10^{-7}$ pb and $m_{\widetilde{\chi}_{1}^{0}}$ = 107 GeV, and the benchmark point in Table~\ref{tab:masses2} at $\sigma_{SI} = 5.4 \times 10^{-10}$ pb and $m_{\widetilde{\chi}_{1}^{0}}$ = 152 GeV. The constraints from previous ZEPLIN~\cite{Lebedenko:2008gb}, XENON~\cite{Angle:2007uj}, and CDMS~\cite{Ahmed:2008eu} experiments are also delineated on the plot, in addition to the forthcoming LUX experiment~\cite{LUX}. Despite the fact the WIMP-nucleon cross-section of our Table~\ref{tab:masses1} benchmark point in Fig.~\ref{fig:sigma_plot} is above the CDMS II upper limit, experimental uncertainties and QCD corrections can account for this variation. On the contrary, the entire region of the experimentally allowed parameter space with large $M_{5}$ and small $m_{0}$ has WIMP-nucleon cross-sections less than the recent CDMS II upper limits. The present model also possesses regions of the parameter space where the neutralino can account for only a small portion of the overall composition of the total observed dark matter. The remaining fraction of the observed relic density in this situation would be composed of other particles, such as axions or cryptons, or additional astrophysical matter. In order to enable a direct comparison of Fig.~\ref{fig:sigma_plot} to the data from CDMS II and other direct-detection experiments, for those points in Fig.~\ref{fig:sigma_plot} with a relic density less than the observed WMAP 2$\sigma$ data, we plot a modified WIMP-nucleon cross-section $\sigma_{SI} \times \frac{\Omega_{\chi}}{\Omega_{WMAP}}$.

\begin{figure}[t]
	\centering
		\includegraphics[width=0.49\textwidth]{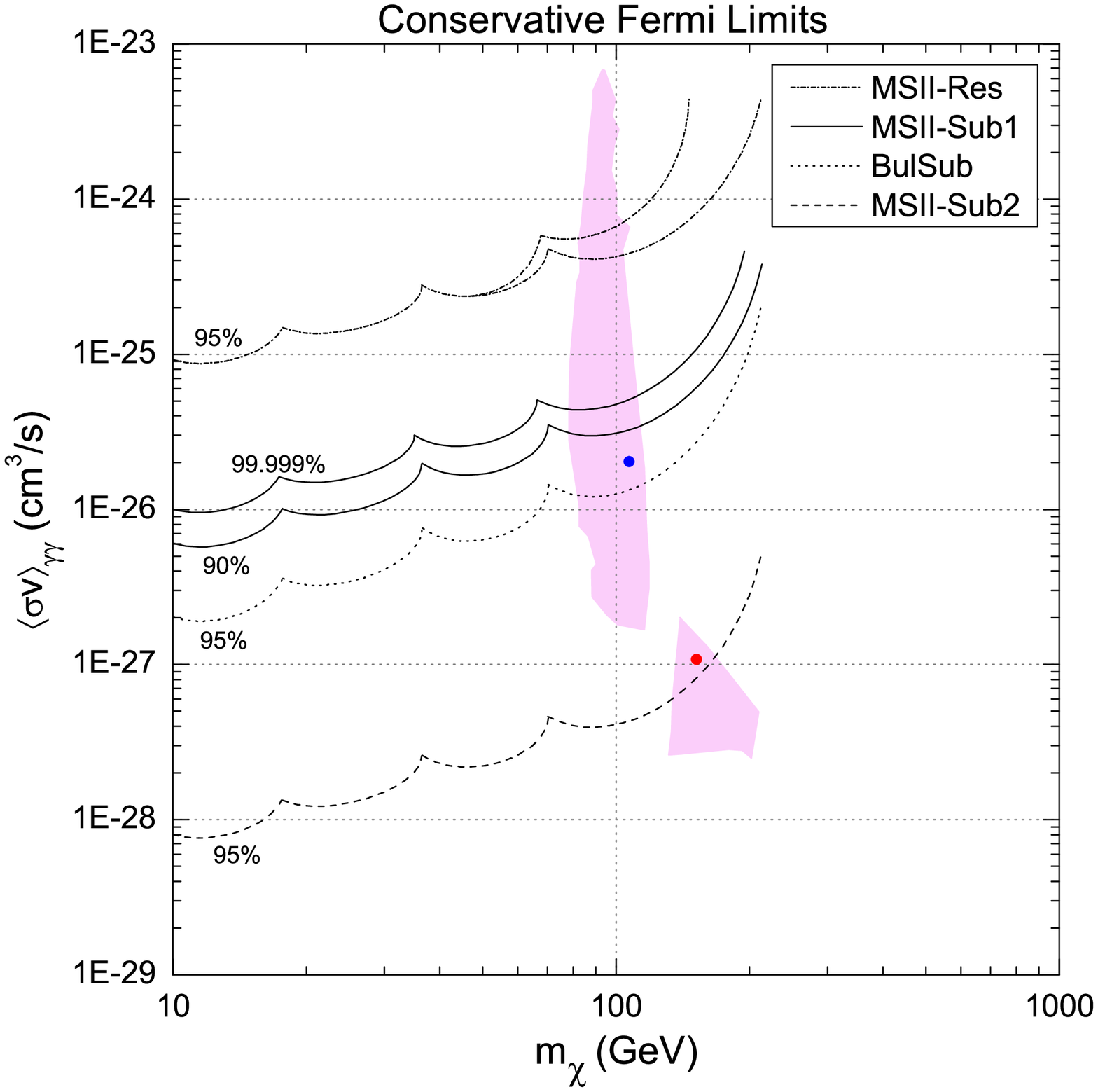}
		\includegraphics[width=0.49\textwidth]{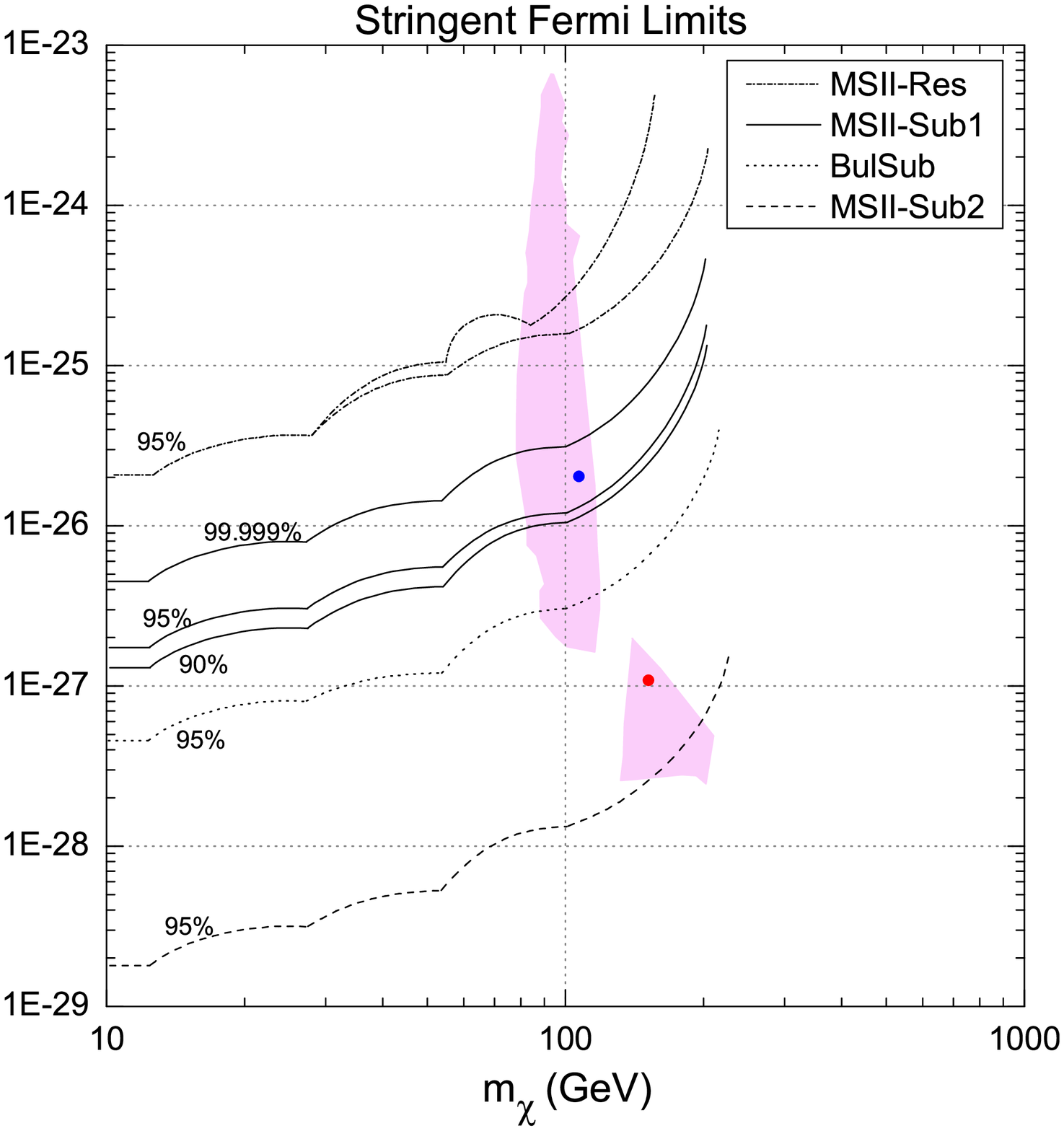}
		\caption{${\cal F}$-$SU(5)$ annihilation cross-section $\left\langle \sigma v \right\rangle_{\gamma\gamma}$ of two neutralinos into two gamma-rays for $\tan \beta = 51$, $A_{0} = m_{0}$, overlaid with the most recent Fermi-LAT constraints. The shaded regions satisfy all experimental constraints. The benchmark point in Table~\ref{tab:masses1} is annotated by the blue points, while the benchmark point in Table~\ref{tab:masses2} is annotated by the red points. For detailed explanations of the four dark matter scenarios (MSII-Res, MSII-Sub1, MSII-Sub2, BulSub), upper limits (conservative, stringent), and confidence limits (90\%, 95\%, 99.999\%), we refer the reader to~\cite{Abdo:2010dk}.}
	\label{fig:fermi_plot}
\end{figure}

Indirect detection experiments search for high energy neutrinos, gamma-rays, positrons, and anti-protons emanating from neutralino annihilation in the galactic halo and core, or in the case of neutrinos, in the core of the sun or the earth. Here, we focus on the annihilation cross-section $\left\langle \sigma v \right\rangle_{\gamma\gamma}$ of two neutralinos into two gamma-rays in the galactic core or halo. Two possible decay channels where WIMPs can produce gamma-rays in the galactic core and halo are $\widetilde{\chi}_{1}^{0} \widetilde{\chi}^{0}_{1} \rightarrow \gamma \gamma$ and $\widetilde{\chi}_{1}^{0} \widetilde{\chi}_{1}^{0} \rightarrow q \overline{q} \rightarrow \pi^{0} \rightarrow \gamma\gamma$. One such current experiment to measure the debris from WIMP annihilations is Fermi-LAT (formerly GLAST)~\cite{Morselli:2002nw}, with new constraints on cross-sections of neutralino annihilations into two gamma-rays~\cite{Abdo:2010dk}. Figure~\ref{fig:fermi_plot} shows that the Fermi-LAT sensitivity has reached the ${\cal F}$-$SU(5)$ parameter space. The ${\cal F}$-$SU(5)$ cross-sections $\left\langle \sigma v \right\rangle_{\gamma\gamma}$ in Figure~\ref{fig:fermi_plot} are calculated using the modified {\tt MicrOMEGAs 2.1} code to include the effects of the 1 TeV vector multiplet in {\tt SuSpect 2.34}. The Fermi-LAT collaboration applies four dark matter structure evolution scenarios (MSII-Res, MSII-Sub1, MSII-Sub2 and BulSub), which we overlay onto the ${\cal F}$-$SU(5)$ parameter space in Figure~\ref{fig:fermi_plot}, as well as conservative and stringent upper limits on $\left\langle \sigma v \right\rangle$. For thorough descriptions of these four dark matter scenarios and upper limits, we refer the reader to~\cite{Abdo:2010dk}. Additionally, the upper 95\% confidence limits for each of the four dark matter scenarios as determined by the Fermi-LAT collaboration are identified in Figure~\ref{fig:fermi_plot}, including the 90\% and 99.999\% confidence limits for the Fermi-LAT reference model MSII-Sub1. For the benchmark point in Table~\ref{tab:masses1}, $\left\langle \sigma v \right\rangle_{\gamma\gamma} = 2.0 \times 10^{-26} ~cm^{3}/s$, in the very near proximity of the Fermi-LAT reference model MSII-Sub1 upper limits for both the conservative and stringent cases, whilst for the benchmark point in Table~\ref{tab:masses2}, $\left\langle \sigma v \right\rangle_{\gamma\gamma} = 1.1 \times 10^{-27} ~cm^{3}/s$, well below the Fermi-LAT reference model MSII-Sub1 upper limits. 

\begin{figure}[ht]
	\centering
		\includegraphics[width=0.75\textwidth]{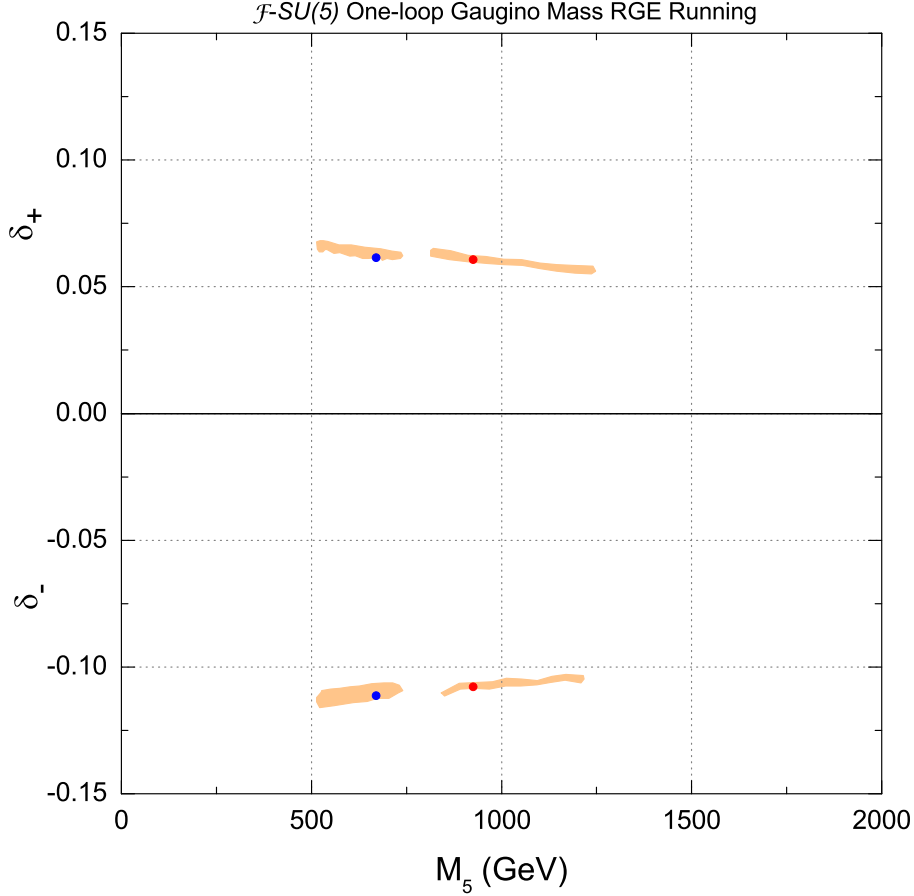}
		\caption{The $\delta_{+}$ and $\delta_{-}$ parameters 
for $\tan \beta = 51$, $A_{0} = m_{0}$ for ${\cal F}$-$SU(5)$. The shaded regions satisfy all experimental constraints. The ${\cal F}$-$SU(5)$ benchmark point in Table~\ref{tab:masses1} is annotated by the blue points, while the benchmark point in Table~\ref{tab:masses2} is annotated by the red points. For the benchmark point in Table~\ref{tab:masses1}, we find $\delta_+$ = 0.0615 and $\delta_-$ = -0.1113, and likewise, for the benchmark point in Table~\ref{tab:masses2}, $\delta_+$ = 0.0607 and $\delta_-$ = -0.1077.}
	\label{FSU(5):delta_plot}
\end{figure}

In the preceding section, we introduced the $\delta_+$ and $\delta_-$ parameters of Eq.~(\ref{delta}) in order to quantify the small two-loop deviations from the mSUGRA gaugino mass relations in Eq.~(\ref{mSUGRA}) at the electroweak scale. The gaugino masses can be measured at the LHC and ILC~\cite{Cho:2007qv, Barger:1999tn}, allowing for a test of these gaugino mass relations. In Fig.~\ref{FSU(5):delta_plot} we present
the deviations $\delta_+$ and $\delta_-$ in  ${\cal F}$-$SU(5)$. Fig.~\ref{FSU(5):delta_plot} demonstrates that $\delta_+$ and $\delta_-$ are indeed small, with $\delta_+ \approx 6\%$ and $\delta_- \approx -11\%$ for ${\cal F}$-$SU(5)$. 
However, these deviations are larger than the F-$SU(5)$ second loop deviations 
in Ref.~\cite{Li:2010mr}. Thus, it is imperative we understand the reason for the larger
deviations in ${\cal F}$-$SU(5)$. 

In an effort to compare the deviations in ${\cal F}$-$SU(5)$, we calculate the deviations in mSUGRA,
{\it i.e.}, the minimal $SU(5)$ model with gravity mediated supersymmetry breaking.
We consider two cases: (i) Two-loop RGE running for the SM gauge couplings and one-loop RGE
running for the gaugino masses; (ii) Two-loop RGE running for both the gauge couplings
and the gaugino masses. We find that in case (i), the order of $M_i/\alpha_i$ is the
same as that in ${\cal F}$-$SU(5)$ given in Eq.~(\ref{FSUV-GM}), while in case (ii)
we have
\begin{eqnarray}
\left({\frac{M_i}{\alpha_i}}\right)_{\rm L} ~=~{\frac{M_3}{\alpha_3}}
~,~~~
\left({\frac{M_i}{\alpha_i}}\right)_{\rm M} ~=~{\frac{M_1}{\alpha_1}}
~,~~~
\left({\frac{M_i}{\alpha_i}}\right)_{\rm S} ~=~{\frac{M_2}{\alpha_2}}
~.~\,
\end{eqnarray}

\begin{figure}[ht]
	\centering
		\includegraphics[width=0.49\textwidth]{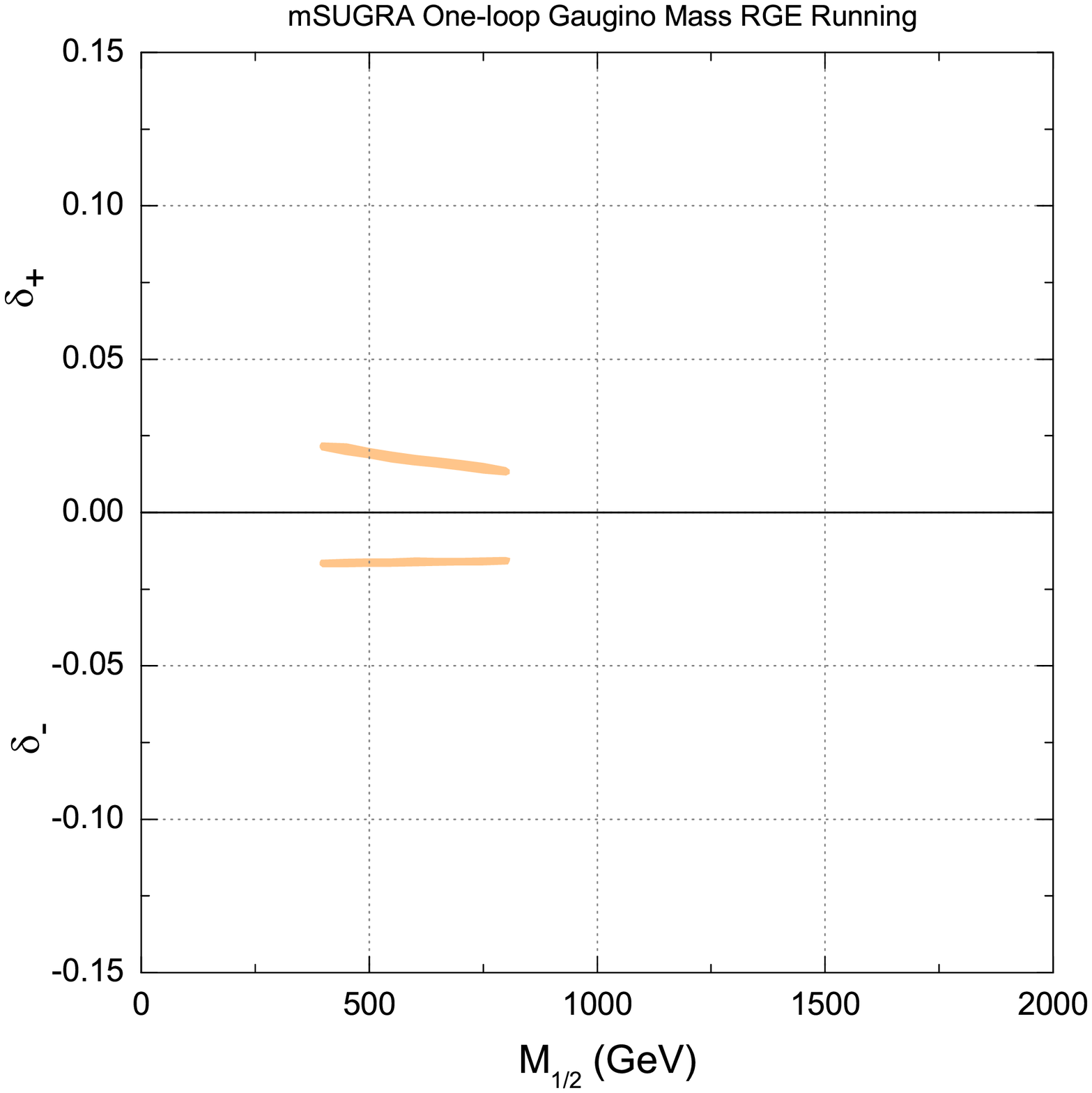}
		\includegraphics[width=0.49\textwidth]{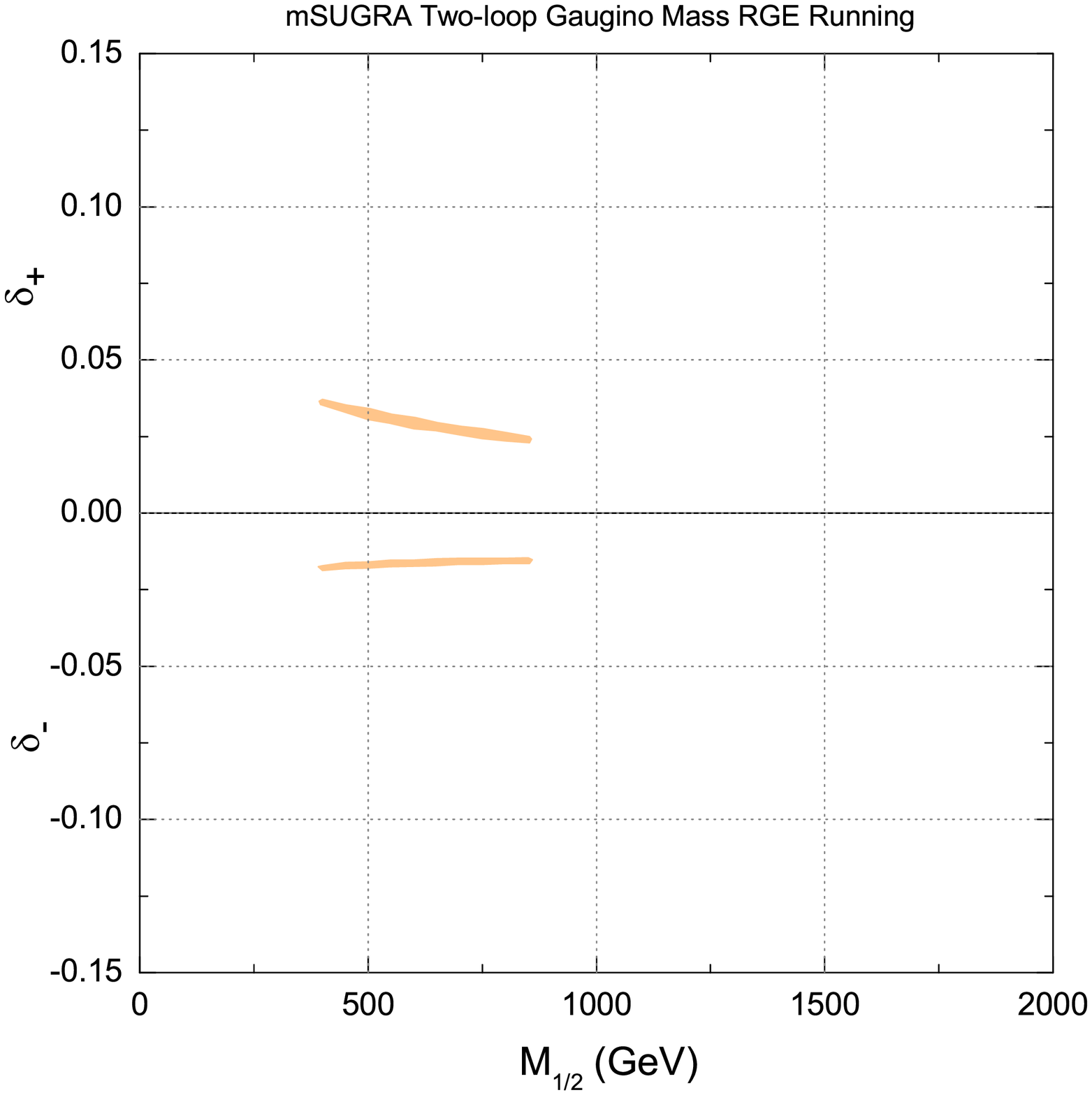}
		\caption{The $\delta_{+}$ and $\delta_{-}$ parameters for $\tan \beta = 51$, $A_{0} = m_{0}$ for case (i)  (left) and case (ii) (right). The shaded regions satisfy all experimental constraints.}
	\label{mSUGRA:delta_plot}
\end{figure}

We present the small deviations $\delta_+$ and $\delta_-$ in mSUGRA for both
case (i) and case (ii) in Fig.~\ref{mSUGRA:delta_plot}.
We find that for case (i) we have  $\delta_+ \approx 1.8\%$ and  
$\delta_+ \approx -1.6\%$, and for case (ii) we have 
 $\delta_+ \approx 3\%$ and $\delta_- \approx -1.6\%$. These deviations are
similar to the  F-$SU(5)$ second loop deviations 
in Ref.~\cite{Li:2010mr}, and are indeed smaller than these in 
${\cal F}$-$SU(5)$. Therefore, we conclude that the existence of
TeV scale vector-like particles in ${\cal F}$-$SU(5)$ enlarge the deviations, 
thereby presenting a potential opportunity to experimentally infer the underlying theory at high energies 
since there is a rather compelling need for TeV scale vector-like particles 
in ${\cal F}$-$SU(5)$. On the contrary, unflipped GUTs require no such 
TeV scale vector-like particles. The vital point here is that an experimental 
measurement of the two-loop deviations through the $\delta_{\pm}$ parameters can 
certainly assist us in determining whether the underlying theory at 
the string scale is a flipped or unflipped GUT.



There is a noticeable resemblance of the experimentally allowed parameter space in Fig.~\ref{fig:movsm5_plot} with that of mSUGRA. For comparison, we study a recent analysis of the mSUGRA parameter space in~\cite{Allahverdi:2009sb}. Upon closer examination of the two models, we see that it appears the vector-like particles and partial gaugino mass universality are shifting the upper narrow WMAP strip to a larger gaugino mass and smaller $m_{0}$. The essential aspect is that the gaugino mass RGE running is changed significantly due to the additional vector-like particles. The consequences of this include a heavier LSP for large $m_{0}$ when compared to mSUGRA, with the LSP mass increasing to a limited range centered around 100 GeV. Recall, our benchmark point from the WMAP strip at large $m_{0}$ is $m_{\widetilde{\chi}_{1}^{0}}$ = 107 GeV. In contrast, the LSP mass in the experimentally allowed region of mSUGRA for large $m_{0}$ and small $m_{1/2}$ is well below 100 GeV. Moreover, the experimentally allowed region of a larger gaugino mass and smaller $m_{0}$ gets shifted to a heavier gaugino mass. In these regions of small $m_{0}$, as already noted, the WMAP relic density is generated through stau-neutralino coannihilation. However, as we described, in ${\cal F}$-$SU(5)$ this region of large $M_{5}$ and small $m_{0}$ possesses WIMP-nucleon cross-sections well below the CDMS II upper limit.

It is interesting to consider what we could see at the LHC for the  ${\cal F}$-$SU(5)$ model framework. We calculate the differential cross-sections and branching ratios with {\tt PYTHIA 6.411}~\cite{Sjostrand:2006za}, using the revised {\tt SuSpect 2.34} code that includes the effects of the vector-like particle contributions to compute the sparticle masses for input into {\tt PYTHIA 6.411}. The three lightest sparticles for the $\tan \beta = 51$ Table~\ref{tab:masses1} benchmark point are $\widetilde{\chi}_{1}^{0} < \widetilde{\chi}_{1}^{\pm} < \widetilde{\chi}_{2}^{0}$. The production of gluinos $\widetilde{g}$ and squarks $\widetilde{q}$ have the largest differential cross-sections at the LHC.  The $\widetilde{q}$ decays to a gluino and hadronic jet due to the noticeably lighter mass of the gluino in comparison to the squarks, while the $\widetilde{g}$ will produce a neutralino or chargino along with a hadronic jet. The $\widetilde{\chi}_{2}^{0}$ produce $Z_{0}$ through $\widetilde{\chi}_{2}^{0} \rightarrow Z_{0} \widetilde{\chi}_{1}^{0}$ with a branching ratio of 64\%. We will get $b$ quarks from the remainder of the $\widetilde{\chi}_{2}^{0}$ through light Higgs in the process $\widetilde{\chi}_{2}^{0} \rightarrow h_{0} \widetilde{\chi}_{1}^{0}$ with a branching ratio of 36\%. These light Higgs will in turn decay to $b \overline{b}$ with a 75\% branching ratio. Leptons and hadronic jets will result from the decay $\widetilde{\chi}_{1}^{\pm} \rightarrow W^{\pm} \widetilde{\chi}_{1}^{0}$, where this is the only kinematically allowed $\widetilde{\chi}_{1}^{\pm}$ process. Thus, this benchmark point will produce mainly $W$ and $Z$ bosons, and some $b$ quarks through light Higgs $h_{0}$ at LHC. Additionally, the vector-like particles can couple to SM fermions and Higgs fields via Yukawa interactions, and they can then decay to SM fermions and Higgs fields, as well as the LSP neutralino.

We see quite different LHC states for the benchmark point in Table~\ref{tab:masses2}. This point resides in the region of the experimentally allowed parameter space that generates the WMAP relic density through stau-neutralino coannihilation. Hence, the four lightest sparticles for this benchmark point are $\widetilde{\chi}_{1}^{0} < \widetilde{\tau}_{1}^{\pm} < \widetilde{\chi}_{2}^{0} \sim \widetilde{\chi}_{1}^{\pm}$. Here, as in the benchmark point in Table~\ref{tab:masses1}, the gluino is lighter than the squarks, so all squarks will predominantly decay to a gluino and hadronic jet, with a small percentage of squarks producing a neutralino or chargino, plus a jet, and the gluinos will then decay to neutralinos or charginos as well. The result is low-energy tau through the processes $\widetilde{\chi}_{2}^{0} \rightarrow \widetilde{\tau}_{1}^{\mp} \tau^{\pm} \rightarrow \tau^{\mp}\tau^{\pm} \widetilde{\chi}_{1}^{0}$ and $\widetilde{\chi}_{1}^{\pm} \rightarrow \widetilde{\tau}_{1}^{\pm} \nu_{\tau} \rightarrow \tau^{\pm}\nu_{\tau} \widetilde{\chi}_{1}^{0}$. The LHC final states of low-energy tau in the ${\cal F}$-$SU(5)$ stau-neutralino coannihilation region are similar to those same low-energy LHC final states in mSUGRA, however, in the stau-neutralino coannihilation region of mSUGRA, the gluino is typically heavier than the squarks. Again, the strong coupling effects from the additional vector-like particles on the gaugino mass RGE running reduce the physical gluino mass below the squark masses in ${\cal F}$-$SU(5)$. As a consequence, the LHC final low-energy tau states in the stau-neutralino coannihilation regions of ${\cal F}$-$SU(5)$ and mSUGRA will differ in that in ${\cal F}$-$SU(5)$, the low-energy tau states will result largely from neutralinos and charginos produced by gluinos, as opposed to the low-energy tau states in mSUGRA resulting primarily from neutralinos and charginos produced from squarks. 

The first year LHC run in 2011 at 7 TeV and anticipated luminosity of 1-2 $fb^{-1}$ is expected to reach gluino production possibly up to 1 TeV, which makes the light gluino in our study here tantalizing for discovery in the early LHC run. However, for our tan$\beta$ = 51 benchmark study in this work, the heavy squarks are most likely too heavy for production in this early LHC phase. Nonetheless, a closer examination of smaller values of tan$\beta$ shows correspondingly lighter spectra, indicating that signals for tan$\beta \leq 16$ are within the first year LHC discovery reach. Additionally, in contrast to our tan$\beta$ = 51 points studied here where the gluino is lighter than all the squarks, for tan$\beta \leq 20$ spectra, the lightest stop is the lone squark lighter than the gluino, providing the potential for a very unique signal. This same $\widetilde{t} < \widetilde{g} < \widetilde{q}$ mass pattern has also been uncovered in follow-up studies where this model has been extended to include the $SU(5)\times U(1)_X$ unification~\cite{Li:2010ws,Li:2010mi,Li:2011dw}. We shall study this signal much more in-depth in our future work.



\section{Proton Decay}

Instability of the proton is an essential signature of GUTs,
the merger of the SM forces necessarily linking quarks to leptons,
and providing a narrow channel $p \!\rightarrow\! {(e\vert\mu)}^{\!+}\! \pi^0$
of dimension six decay via heavy gauge boson exchange.
We have undertaken a comprehensive analysis of the proton decay rate for the available
parameter space of the ${\cal F}$-$SU(5)$ model of the current report, employing the
methods established in Refs.~\cite{Li:2009fq, Li:2010dp} and the further citations therein.

The 50-kiloton (kt) water \v{C}erenkov detector of the Super-Kamiokande facility
has set lower bounds of $8.2\times 10^{33}$ and $6.6\times 10^{33}$ years
at the $90\%$ confidence level for the partial lifetimes in the
$p\rightarrow e^+ \pi^0$ and $p\rightarrow \mu^+ \pi^0$ modes~\cite{:2009gd}.
Hyper-Kamiokande is a proposed 1-Megaton detector, about 20 times larger
volumetrically than Super-K~\cite{Nakamura:2003hk},
which we can expect to explore partial lifetimes 
up to a level near $2\times 10^{35}$ years for
$p \!\rightarrow\! {(e\vert\mu)}^{\!+}\! \pi^0$ across a decade long run.
The proposal for the DUSEL experiment~\cite{Raby:2008pd,cline}
features both water \v{C}erenkov and liquid Argon (which is around five times
more sensitive per kilogram to $p\rightarrow K^+ {\bar \nu}_{\mu}$ than water)
detectors, in the neighborhood of 500 and 100 kt respectively,
with the stated goal of probing partial lifetimes into the
order of $10^{35}$ years for both the neutral pion and $K^+$ channels.
Flipped $SU(5)$ evades the dangerously rapid $p\rightarrow K^+ {\bar \nu}$
dimension five triplet Higgsino mediated decay
by way of the missing-partner mechanism~\cite{AEHN-0}, which
realizes Higgs doublet-triplet splitting
naturally, without the side effect of strong triplet mixing. 

Proton lifetime in the $p \!\rightarrow\! {(e\vert\mu)}^{\!+}\! \pi^0$
dimension six mode is proportional in the fourth power to the GUT mass scale,
and inversely proportional in the fourth power to the GUT coupling.
This extreme sensitivity argues for great care in the
selection and study of a unification scenario.
The F-theory derived GUT models which feature
additional vector-like multiplets at the TeV scale
modify the renormalization group to yield a substantial
increase in the $SU(3)_C\times SU(2)_L$ unified coupling,
which translates in turn into something like a seven-fold
enhancement of the proton decay rate.
Moreover, inclusion of TeV scale vector multiplets in the renormalization significantly magnifies
the separation of the flipped scale $M_{32}$, at which $SU(5)$ breaks and proton decay is
established, from the point $M_{51}$ of true Grand Unification,
which can be extended to the order of the reduced Planck mass.

In our calculations, we take the sparticle and Higgs spectrum
generated by {\tt SuSpect 2.34}, modified for flipped
$g_2 = g_3$ unification, as input for the light threshold corrections.  These are applied
individually to each of the three gauge couplings, taking into account the distinct
quantum numbers of each field.  The second loop contribution is likewise individually
numerically determined for each gauge coupling, including the top and bottom quark Yukawa couplings
from the third generation, taken themselves in the first loop.  All three gauge couplings are 
integrated recursively with the second loop into the Yukawa coupling renormalization, with
the boundary conditions at $M_{\rm Z}$ set by the value of $\tan \beta = 44$.  We take
this value, reduced modestly from that of $51$ employed elsewhere in this report, to avoid
pathologies in convergence of the renormalization.  We emphasize that this slightly modified value of tan$\beta$ = 44 is employed only for the proton lifetime calculations, whereas the value of tan$\beta$ = 51 is used in all other calculations for the benchmark points. Finally, recognizing that
the second loop itself influences the upper limit $M_{32}$
of its own integrated contribution, this feedback is accounted for in the dynamic
calculation of the unification scale.
The benchmark spectra of the selected points featured in
Tables~\ref{tab:masses1} and~\ref{tab:masses2} yield central
lifetimes of $2.8\times10^{34}$ and $6.0\times10^{34}$ years, respectively. 

In addition to the light $M_Z$-scale threshold corrections from the superpartner's entry into the RGEs,
there may also be shifts occurring near the
$M_{23}$ unification point due to heavy Higgs fields and the
broken gauge generators of $SU(5)$.
The light fields carry strong correlations to cosmology and low energy
phenomenology, as is indeed the central theme of this work, so that we are guided
toward plausible estimates of their mass distribution.
The heavy thresholds are somewhat more difficult to pin down, although 
we may conclusively suggest that they act only to lengthen, rather than reduce,
the dimension six proton lifetime~\cite{Li:2010dp}.

Including uncertainties for parameter input and light and heavy
thresholds, we conclude that the central values for proton decay,
and also a large fraction of the plausible variation,
are indeed within the reach of proposed
next-generation experiments such as Hyper-Kamiokande and DUSEL.
Additionally, detectability of TeV scale vector supermultiplets
at the LHC presents an opportunity for cross correlation of results
between the most exciting particle physics experiments of the coming decade.

We emphasize that the cumulative effect of flipped $SU(5)$ grand unification in the
F-theory model building context, applying the freshly detailed methods of analysis
recently described, and focusing on the phenomenologically preferred parameter space,
is surprisingly fast proton decay.



\section{Conclusion}

We have considered gravity mediated supersymmetry breaking in ${\cal F}$-$SU(5)$. The gaugino masses are not unified at the traditional grand unification scale, though we do indeed obtain the mSUGRA one-loop gaugino mass relation at the electroweak scale. However, the gaugino mass relation will have a small two-loop deviation, and this deviation may be measurable at the LHC and ILC. There is a considerable need for TeV scale vector-like particles in ${\cal F}$-$SU(5)$, while unflipped GUTs, such as mSUGRA or F-$SU(5)$, require no such vector-like particles. In light of this key distinction between flipped and unflipped GUTs, we introduced a parameter to measure the small two-loop deviation from the mSUGRA gaugino mass relation at the electroweak scale.

To implement a numerical analysis, we modified a popular and well-established public RGE code to incorporate the effects of TeV scale vector-like particles. In this work, we employed two-loop RGEs for the gauge couplings, but only considered one-loop RGEs for the Yukawa couplings and soft-supersymmetry breaking terms, though we look to extend this to all two-loop supersymmetry breaking RGEs in our future work. The results lead us to conclude there is a clear disparity in the extent of the deviation between GUTs with vector-like particles, such as ${\cal F}$-$SU(5)$, and GUTs without vector-like particles, such as mSUGRA or F-$SU(5)$.
The predicted correlation between largeness of the deviation from the mSUGRA gaugino mass relation
and the existence of vector-like particles can be tested at the colliders.

Furthermore, we have determined the viable parameter space of this model which simultaneously satisfies all the current experimental constraints and is consistent with the findings of CDMS II. The cross-section of two neutralinos into two gamma-rays for the experimentally allowed regions of the parameter space was computed and assessed against the first published Fermi-LAT measurement. The results showed the ${\cal F}$-$SU(5)$ parameter space is consistent with the recent Fermi-LAT findings. Finally, we have calculated the proton lifetime for these experimentally allowed regions, and found it to be within the reach of the future Hyper-Kamiokande and DUSEL experiments.

A wealth of experimental data is on the horizon, so it is imperative that phenomenologically appealing GUTs, for instance ${\cal F}$-$SU(5)$, be researched so that unambiguous experimental predictions may be presented. These predictions will be key milestones in deducing the underlying theory at high energies as we progress through the next few exciting years of LHC, direct dark matter detection, Fermi-LAT, and proton decay experiments. We have supplemented the conventional bottom-up analysis of traditional GUTs to include TeV scale vector-like particles, and our results feature encouraging prospects for the experimental determination of whether high-energy theory indeed admits these proposed multiplets. We believe that in the next few years experiment will certainly have something key to say about ${\cal F}$-$SU(5)$ in particular and string theory in general.


\begin{acknowledgments}


This research was supported in part 
by  the DOE grant DE-FG03-95-Er-40917,
by the Natural Science Foundation of China 
under grant No. 10821504 (TL),
and by the Mitchell-Heep Chair in High Energy Physics.

\end{acknowledgments}


\end{document}